\documentclass[letterpaper]{jpconf}
 \usepackage{epsfig}
\usepackage{cite}
 \usepackage{setspace}
 \usepackage{latexsym}

\begin{document}

\def\Vus{\mathrm{|V_{us}|}}
\def\BmuBe{\frac{B(\tau^-\rightarrow\mu^-\bar{\nu_{\mu}}\nu_{\tau})}{B(\tau^-\rightarrow e^-\bar{\nu_{e}}\nu_{\tau})}}
\def\BpiBmu{\frac{B(\tau^-\rightarrow\pi^-\nu_{\tau})}{B(\pi^-\rightarrow\mu^-\bar{\nu_{\mu}})}}
\def\BkBmu{\frac{B(\tau^-\rightarrow K^-\nu_{\tau})}{B(K^-\rightarrow\mu^-\bar{\nu_{\mu}})}}
\def\gmuge{\left(\frac{g_{\mu}}{g_e}\right)^2 = \BmuBe\frac{f(m_e^2/m_\tau^2)}{f(m_\mu^2/m_\tau^2)}}
\def\gtaugmuPi{\left(\frac{g_{\tau}}{g_\mu}\right)^2 = \BpiBmu\frac{2m_\pi m_\mu^2\tau_{\pi}}{\delta_{\tau/\pi}m_\tau^3\tau_{\tau}}\left(\frac{1-m_\mu^2/m_\pi^2}{1-m_\pi^2/m_\tau^2}\right)^2}
\def\gtaugmuK{\left(\frac{g_{\tau}}{g_\mu}\right)^2 = \BkBmu\frac{2m_K m_\mu^2\tau_{K}}{\delta_{\tau/K}m_\tau^3\tau_{\tau}}\left(\frac{1-m_\mu^2/m_K^2}{1-m_K^2/m_\tau^2}\right)^2}

\def\SCCOmega{\tau^-\rightarrow \omega\pi^-\nu_{\tau}}
\def\SCCEta{\tau^-\rightarrow \eta'(958)\pi^-\nu_{\tau}}

\title{Recent Results From BaBar in $\tau$ Physics}

\author{Mateusz Lewczuk [BaBar Collaboration]}

\address{Dept. of Physics and Astronomy, University of Victoria, P.O. Box 3055, Victoria, B.C., V8W 3P6, Canada}

\section{Abstract}

The BaBar collaboration has accumulated over 400 million $\tau$-pairs which can be used to study charged leptonic and hadronic weak currents to unprecedented precision. This note presents results on lepton universality, measurements of $\Vus$, and searches for $\tau $ decays which violate lepton flavour conservation, or $\tau$ decays that proceed through a suppressed second class current. 

\section{Lepton Universality}

Under lepton universality it is assumed that the charged weak coupling between the leptons and the gauge bosons is the same for all generations of leptons. The $\tau-\mu$ universality is determined with \[\gtaugmuPi,\] \[\gtaugmuK,\] where the B($\tau^-\rightarrow\pi^-\nu_{\tau}$) and B($\tau^-\rightarrow K^-\nu_{\tau}$) branching fractions are measured by BaBar, $m_\tau$ is the $\tau$ mass $\tau_\tau$ is the $\tau$ lifetime $\tau_\pi/\tau_K$ and $m_\pi/m_K$ are the $\pi/K$ lifetimes and masses, respectively, and $\delta_{\tau/\pi}$ and $\delta_{\tau/K}$ are radiative corrections \cite{LepRadCorr}. BaBar measures $|\frac{g_{\tau}}{g_\mu}|$ = $0.9859\pm0.0057$ and $|\frac{g_{\tau}}{g_\mu}|$ = $0.9836\pm0.0087$ using $\tau^-\rightarrow\pi^-\nu_{\tau}$, and $\tau^-\rightarrow K^-\nu_{\tau}$ respectively. 

Similarly, the $\mu-e$ universality is determined with \[\gmuge\]
where $\frac{B(\tau^-\rightarrow\mu^-\bar{\nu_{\mu}}\nu_{\tau})}{B(\tau^-\rightarrow e^-\bar{\nu_{e}}\nu_{\tau})} = 9.796\pm0.016\pm0.035$\cite{chconj} is measured by BaBar and $f(x)=1-8x+8x^3-x^4-12x^2logx$, where the mass of the neutrino is assumed to be zero. The ratio $|\frac{g_{\mu}}{g_e}|$ is measured to be $1.0036\pm0.0020$ which is consistent with the previous world average $1.0000\pm0.0020$\cite{ref1}. It is possible to use this result to determine the total hadronic branching fraction of the $\tau$, $B_{\tau,had}=(64.823\pm0.059)\%$, assuming $\mu-e$ universality \cite{Banerjee:2008hg}.

\section{$\mathrm{|V_{us}|}$}

The unitarity of the CKM matrix requires, $\mathrm{V_{ud}^2+V_{us}^2+V_{ub}^2=1}$, where $\mathrm{V_{ud}}$, $\mathrm{V_{us}}$, and $\mathrm{V_{ub}}$ are the mixing amplitudes between the up quark and the down, strange and bottom quarks, respectively. From this equation its is possible to predict $\mathrm{V_{us}}=\mathrm{\sqrt{1-|V_{ud}|^2-|V_{ub}|^2}}=0.2262\pm0.0011$ using $\mathrm{V_{ub}}=0.00359\pm0.00016$ and $\mathrm{V_{ud}}=0.97408\pm0.00026$ \cite{Eronen:2007qc}, which is a recent value determined from super allowed $\beta$ decays that is consistent with, and more precise than, the PDG world average. 

\subsection{Determining $\mathrm{V_{us}}$ from inclusive strange $\tau$ decays}

\begin{table}
\centering
\begin{tabular}{|l|l|}
\hline
 Decay Mode & Branching Fraction $(\%)$\\
\hline
$K^-\nu_{\tau}$ & $0.69\pm0.010$ \\
$K^-\pi^0\nu_{\tau}$ & $0.426\pm0.016$\cite{ref2}\\
$\bar{K}^0\pi^-\nu_{\tau}$ & $0.835\pm0.022$\cite{ref3,ref4}\\
$K^-\pi^0\pi^0\nu_{\tau}$ & $0.058\pm0.024$\cite{ref5}\\
$\bar{K}^0\pi^0\pi^-\nu_{\tau}$ & $0.360\pm0.040$\cite{ref5}\\
$K^-\pi^-\pi^+\nu_{\tau}$ & $0.290\pm0.018$\cite{ref6,ref7}\\
$K^-\eta\nu_{\tau}$ & $0.016\pm0.001$\cite{ref8}\\
$(\bar{K}3\pi\nu_{\tau})^-(estimated)$ & $0.074\pm0.030$\cite{ref9}\\
$K_{1}(1270)\nu_{\tau}\rightarrow K^-\omega\nu_{\tau}$ & $0.067\pm0.021$\cite{ref9}\\
$(\bar{K}4\pi\nu_{\tau})^-(estimated)$ & $0.011\pm0.007$\cite{ref9}\\
$K^{*-}\eta\nu_{\tau}$ & $0.014\pm0.001$\cite{ref8}\\
$K^-\phi\nu_{\tau}$ & $0.0037\pm0.0003$\cite{ref6,ref10}\\
\hline
Total & $2.8447\pm0.0688$\\
\hline
\end{tabular}
\caption{This table displays the branching fractions values for strange $\tau$ decays used in determining the total $\tau$ to strange branching fraction for the inclusive $|V_{us}|$ measurement.}
\label{fig:StrangeTau}
\end{table}

BaBar has recently measured B($\tau^-\rightarrow K^-\pi^0\nu_{\tau}$), B($\tau^-\rightarrow K^-\pi^-\pi^+\nu_{\tau}$), B($\tau^-\rightarrow K^-\phi\nu_{\tau}$), and B($\tau^-\rightarrow \bar{K}^0\pi^-\nu_{\tau}$) (see Table \ref{fig:StrangeTau}). These decays contribute to the total strange branching fraction of the $\tau$. The total strange $\tau$ branching fraction can be used to determine the $\mathrm{|V_{us}|}$ mixing element within the framework of the Operator Product Expansion (OPE), and Finite Energy Sum Rules (FESR) \cite{FESR}. In this approach the matrix element $\mathrm{|V_{us}|}$ is defined according to \[\mathrm{|V_{us}|= \sqrt{ R_{\tau,strange}/\left[\frac{R_{\tau,non-strange}}{|V_{ud}|^2} - \delta R_{\tau,theory}\right] }}, \] where $\delta R_{\tau ,theory}$ is a correction factor that accounts for differences in the strange and non-strange $\tau$ decays, $\mathrm{|V_{ud}|}$ is determined from super allowed $\beta$ decays\cite{Eronen:2007qc}, and $R_{\tau , strange}$, and $R_{\tau , non-strange}$ are the total strange and non-strange branching fraction ratios of the $\tau$ lepton respectively, normalized to the $B(\tau^-\rightarrow e^-\bar{\nu}_e \nu_{\tau})$. The total strange branching fraction of the $\tau$ is determined from the values shown in Table \ref{fig:StrangeTau}, and the non-strange branching fraction is taken from $B_{\tau\rightarrow non-strange} = B_{\tau\rightarrow hadrons} - B_{\tau\rightarrow strange}$ where $B_{\tau\rightarrow hadrons}$ is determined in accordance with lepton universality as discussed in the previous section. Using the above equation and the results in Table \ref{fig:StrangeTau}, we obtain $\mathrm{|V_{us}|}=0.2159\pm0.0030$ which is below the value $\mathrm{|V_{us}|} = 0.2262\pm0.0011$, estimated from CKM unitarity and super allowed $\beta$ decays.

\subsection{Determining $|V_{us}|$ from $B(\tau^-\rightarrow K^-\nu_{\tau})/B(\tau^-\rightarrow\pi^-\nu_{\tau})$}

An alternative method for determining $\mathrm{|V_{us}|}$ from $\tau$ decays is by studying the ratio of $B(\tau^-\rightarrow K^-\nu_{\tau})/B(\tau^-\rightarrow\pi^-\nu_{\tau})$. In this approach $\mathrm{|V_{us}|}$ is extracted from \[\frac{B(\tau^-\rightarrow K^-\nu_{\tau})}{B(\tau^-\rightarrow\pi^-\nu_{\tau})} = \frac{f_K^2|V_{us}|^2(1-\frac{m_K^2}{m_\tau^2})^2}{f_\pi^2|V_{ud}|^2(1-\frac{m_pi^2}{m_\tau^2})^2}\times\frac{\delta_{\tau^-\rightarrow K^-\nu_{\tau}}}{\delta_{\tau^-\rightarrow\pi^-\nu_{\tau}}},\] where $B(\tau^-\rightarrow K^-\nu_{\tau})$ and $B(\tau^-\rightarrow\pi^-\nu_{\tau})$ are measured branching fractions, the ratio of the K and $\pi$ decay constants is $f_K^2/f_\pi^2=1.189\pm0.007$\cite{Follana:2007uv}, $\mathrm{|V_{ud}|}$ is from \cite{Eronen:2007qc}, and $\delta_{\tau^-\rightarrow K^-\nu_{\tau}}/\delta_{\tau^-\rightarrow\pi^-\nu_{\tau}} = 1.0003\pm0.0044$\cite{Banerjee:2008hg}. We obtain $\mathrm{|V_{us}|} = 0.2254\pm0.0023$, which is consistent with prediction.

\section{Lepton Flavour Violation in $\tau$ Decays}

Lepton flavour violating (LFV) decays are expected in the Standard Model (SM) as well as new physics models. Rates for decays that violate lepton flavour in the SM are predicted to be many orders of magnitude lower than rates predicted in various new physics models\cite{LFVSources}. Experimental limits can only access LFV predictions by new physics models, and so a detection of a LFV decay is a sign of new physics.

\begin{table}
\centering
\begin{tabular}{|l|l|}
\hline
 Mode & $UL_{90}$\\
\hline
$\tau^-\rightarrow e^-K^0_s$   & $4.0\times10^{-8}$\\
$\tau^-\rightarrow \mu^-K^0_s$ & $3.3\times10^{-8}$\\
\hline
$\tau^-\rightarrow e^-\omega$  & $1.1\times10^{-7}$\\
$\tau^-\rightarrow \mu^-\omega$& $1.0\times10^{-7}$\\
\hline
\end{tabular}
\caption{This table shows the 90$\%$ confidence level upper limit set by BaBar on the $\tau^-\rightarrow l^-\omega$ ($l = e/\mu$)\cite{Aubert:2007kx}, and the preliminary results on $\tau^-\rightarrow l^-K_s^0$($l = e/\mu$).}
\label{fig:TauTolWlKs}
\end{table}

BaBar searches for LFV $\tau$ decays by looking for signal in a region defined by \[\Delta m = m_{rec} - m_\tau \] \[\Delta E = E_{rec} - \sqrt{s}/2,\] where $m_{rec}$ is the reconstructed mass of the signal channel, $m_{\tau}$ is the $\tau$ mass, $E_{rec}$ is the reconstructed energy of the signal channel and $\sqrt{s}$ is the square root of the centre mass energy of the $e^+e^-$ collision. Since no neutrinos are present, a signal in an LFV decay is expected to peak at $\Delta m$ and $\Delta E$ equaling 0. 

BaBar has recently conducted searches for LFV in $\tau^-\rightarrow l^-l^+l^-$\cite{Aubert:2007pw}, $\tau^-\rightarrow l^- \omega$\cite{Aubert:2007kx}, $\tau^-\rightarrow l^-V^0$ (where $V^0$ is either a $\rho^0$, $K^{*0}$, $\bar{K}^{*0}$, or $\phi$), and $\tau^-\rightarrow l^- K_s^0$ ($l = e/\mu$ for all modes). The results for the $\tau^-\rightarrow l^-l^+l^-$ mode show no signal and set the $90\%$ confidence level upper limits on the branching fractions in the range $4-8\times 10^{-8}$ for the 6 LFV combinations of $e/\mu$ in the final state. Searches for the $\tau^-\rightarrow l^-V^0$ decays set the 90$\%$ confidence level upper limit on the branching fractions in the $1-18\times 10^{-8}$ range. The results for $\tau^-\rightarrow l^-\omega$ and $\tau^-\rightarrow l^-K_s^0$ are shown in Table \ref{fig:TauTolWlKs}. All LFV searches see no signal and new $90\%$ confidence level upper limits are assigned.

\section{Second Class Currents in $\tau$ Decays}

Decays suppressed by G-parity conservation, are said to proceed through a second class current (SCC).  The final states in $\tau$ decays can be classified as first class (favored) and second class (suppressed) according to their G partiy. The decay $\tau^-\rightarrow X^-\nu_{\tau}$, where X$^-$ is an allowed mesonic state, is a first class state if the $\mathrm{J^{PG}}$ value of the decay product is $\mathrm{J^{PG}} =  0^{--},1^{+-},1^{-+}$ and second class if $\mathrm{J^{PG}} =  0^{+-},1^{++}$. 

BaBar has conducted searches for SCC in the $\SCCOmega$ and $\SCCEta$ channels. The former decay has a first class and second class component. The first and second class components are seperated by analyzing the $cos(\theta_{\pi\omega})$ distribution, where $\theta_{\pi\omega}$ is the angle between the spectator $\pi$ and the normal to the $\omega$ decay plane. A data sample of 347$fb^{-1}$, is fitted with \[F(cos\theta_{\pi\omega}) = N[(1-\epsilon)F^{FCC}_{LJ=11}(cos\theta_{\pi\omega}) + \epsilon F^{SCC}_{LJ=01}(cos\theta_{\pi\omega}),]\] where $\epsilon$ is the fraction of the $\SCCOmega$ decays that proceed through a second class current. The fit returns $\epsilon = 5.5\pm5.8^{+0.8}_{-5.5}$, consistent with no second class current contribution. The $90\%$ confidence level upper limit on the ratio of SCC to FCC is set at 0.69$\%$\cite{SCCOmegaRef}.

Unlike the $\SCCOmega$ decay, the $\SCCEta$ decay proceeds entirely through a second class current. A search for this mode was conducted by fitting the $\pi^+\pi^-\eta$ invariant mass in the region of the $\eta'(958)$ meson. The fit to a 384$fb^{-1}$ data sample returned $19\pm13$ signal candidates. The 90$\%$ confidence level upper limit is set at $7.2\times10^{-6}$\cite{Aubert:2008nj}.

\section{Conclusion}

The lepton universality measurements at BaBar are consistent with the values of the world averages. 
An estimate of $\mathrm{|V_{us}|}$ using new $\tau$-decay measurements by BaBar and other experiments is found to be lower than the value obtained assuming CKM unitarity together with a new measurement of $\mathrm{|V_{ud}|}$ from $\beta$-decays. In comparison, $\mathrm{|V_{us}|}$ obtained from new BaBar measurements of $B(\tau^-\rightarrow K^-\nu_{\tau})/B(\tau^-\rightarrow\pi^-\nu_{\tau})$ are consistent with this value. More precise limits are set on LFV modes in the $\tau^-\rightarrow l^-l^+l^-$, $\tau^-\rightarrow l^- \omega$\, $\tau^-\rightarrow l^-V^0$, and $\tau^-\rightarrow l^- K_s^0$ decays. ($l = e/\mu$ for all modes). New upper limits are put on the second class current contribution to $\SCCOmega$, and the $\SCCEta$ decay which proceeds entirely through a second class current.

\vskip 2cm


\begin{thebibliography}{99}
\bibitem{chconj}
Charge conjugation is assumed throughout this paper.
\end{thebibliography}
\end{document}